\documentclass[prb,showpacs,floatfix,twocolumn]{revtex4} 

\usepackage{epsfig} 

\def\ZMG{Zn$_{1-x}$Mg$_x$O} 

\begin{document} 

\title{Local structure evolution in polycrystalline \ZMG\/ 
($0\leq{x}\leq{0.15}$) studied by Raman and by synchrotron x-ray pair 
distribution analysis}

\author{Young-Il Kim, Katharine Page, Andi M. Limarga, David R. Clarke,
and Ram Seshadri} 

\affiliation{Materials Department and Materials Research Laboratory,
University of California, Santa Barbara, CA 93106} 

\date{\today} 

\begin{abstract} 

The local structures of \ZMG\/ alloys have been studied by Raman
spectroscopy and by synchrotron x-ray pair distribution function (PDF)
analysis. Within the solid solution range ($0\leq{x}\leq{0.15}$) of \ZMG,
the wurtzite framework is maintained with Mg homogeneously distributed
throughout the wurtzite lattice.  The $E_2^\mathrm{high}$ Raman line of
\ZMG\/ displays systematic changes in response to the evolution of the 
crystal lattice upon the Mg-substitution. The red-shift and broadening 
of the $E_2^\mathrm{high}$ mode are explained by the expansion of hexagonal 
$ab$-dimensions, and compositional disorder of Zn/Mg, respectively. 
Synchrotron x-ray PDF analyses of \ZMG\/ reveal that the Mg atoms have 
a slightly reduced wurtzite parameter $u$ and more regular tetrahedral 
bond distances than the Zn atoms. For both Zn and Mg, the internal 
tetrahedral geometries are independent of the alloy composition.  

\end{abstract} 


\maketitle 

\section{INTRODUCTION}

Polar semiconductors such as wurtzite ZnO and GaN have great potential
for use in the  polarization-doped field effect transistors (PolFETs) 
which exploit a polarization gradient at the channel layer to attain 
higher mobility and higher concentrations of carriers. The performance 
of PolFETs depends primarily on the  interface quality and the magnitude 
of the polarization gradient at the heterojunction.  
Those goals can be approached by interfacing a polar
semiconductor with its alloy derivative, as exemplified by the
fabrication of a PolFET comprising the  interface of
GaN/Ga$_{1-x}$Al$_x$N ($x=0-0.3$).\cite{Rajan} In designing  ZnO-based
PolFETs, \ZMG\/ alloys can be considered as the sub-layer
component, since Mg-substitution effectively alters the polarization
of ZnO while keeping the lattice dimensions nearly unchanged.\cite{Kim} 

In a previous report,\cite{Kim} we have outlined the prospects for
ZnO/\ZMG\/ heterojunctions for PolFET applications, based on structural
analysis of \ZMG\/ alloys using  synchrotron x-ray diffraction. We showed 
that the ionic polarization can be tuned by $\approx$14\% from ZnO to
Zn$_{0.85}$Mg$_{0.15}$O, despite only  small changes in the cell 
volume ($\approx$0.3\%). Also, it was found that Mg can substitute up to
15\% of total Zn in the wurtzite lattice, without any evidence for 
segregation. While the average crystal structures of \ZMG\/ phases
were  accurately determined in the above study, there still remain the
local structural  details of the solid solutions to be understood.
Therefore in a continuing effort, we have performed Raman studies and 
synchrotron x-ray pair distribution function (PDF) analysis of polycrystalline 
\ZMG\/ samples. Raman spectroscopy is known to be useful for examining 
compositional disorder and/or the presence of strain within semiconductor 
alloys.\cite{Rohmfeld,Franz,Richter,Tiong,Ramkumar,Yang} 
In previous Raman studies on wurtzite type crystals, the peak position and 
shape of the $E_2^\mathrm{high}$ phonon mode have  been used to investigate 
the effects of sample grain size, heterogeneous  components, and
defects.\cite{Richter,Tiong} PDF analysis provides a powerful probe of 
local non-periodic  atomic displacements in the short range, and is a suitable 
complement to the $k$-space refinement technique.\cite{Bill,Qiu} 
This ability is very relevant to the possible distinction of tetrahedral
out-of center displacements of Zn and Mg, and provides for a better comparison
platform with density functional calculations of the crystal structure.

In this study, we provide a complete description of polycrystalline \ZMG\/ 
alloys prepared from crystalline Zn$_{1-x}$Mg$_x$(C$_2$O$_4$)$\cdot$2H$_2$O
precursors. We present results of thermogravimetry, ultraviolet/visible 
diffuse-reflectance spectroscopy, infrared, and Raman spectroscopy in 
addition to examining the distinct tetrahedral geometries of Zn and Mg 
using synchrotron x-ray PDF analyses based on supercell structure models. 

\section{EXPERIMENTAL}

Powder samples of \ZMG\/ ($x$ = 0, 0.05, 0.10, and 0.15) were prepared
using an  oxalate precursor route as described previously.\cite{Kim}
Aqueous solutions of zinc  acetate, magnesium acetate, and oxalic acid
were separately prepared and mixed to  precipitate zinc magnesium
oxalates, which are crystalline, single-phase compounds with Zn$^{2+}$ and 
Mg$^{2+}$ homogeneously mixed at the atomic level. The precipitates were 
thoroughly washed with deionized water and dried at 60$^{\circ}$C for 4\,h 
to produce white powders of Zn$_{1-x}$Mg$_x$(C$_2$O$_4$)$\cdot$2H$_2$O, as
verified by powder x-ray diffraction. Subsequently the oxalate dihydrates 
were transformed to \ZMG\/ by heating in air at
550$^{\circ}$C for 24\,h.  The thermal  decomposition of
Zn$_{1-x}$Mg$_x$(C$_2$O$_4$)$\cdot$2H$_2$O was monitored by 
thermogravimetry using a Cahn ThermMax 400 thermogravimetric analyzer
(Thermo  Scientific). For each composition, $\approx$60\,mg of powder
was  heated in air up to 1000$^{\circ}$C at
5$^{\circ}$C\,min$^{-1}$.  Fourier-transform infrared (FT-IR) spectra of
\ZMG\/ powders were recorded in  KBr using a Nicolet Magna 850 FT-IR
spectrophotometer in the transmission mode.  Diffuse-reflectance
absorption spectra were measured for \ZMG\/ in the wavelength  range of
220$-$800\,nm using a Shimadzu UV-3600 spectrophotometer equipped with an 
ISR-3100 integrating sphere. The powder samples were mounted to have flat
surfaces  and $\approx$1.5\,mm thickness. The optical band gap was
determined by extrapolating  the linear part of absorption edge to
zero-absorption level.  Raman measurements were conducted at room
temperature using an optical microprobe  fitted with a single
monochromator (Jobin-Yvon, T64000). Powder samples were compacted
on frosted glass plates and spectra were recorded  in 
backscattering geometry using a 488.08\,nm Ar$^+$ laser with a beam
power  of 50\,mW and a spot size of $\approx$2\,$\mu$m. For each sample,
five acquisitions of 30\,s exposure were performed and the averaged
spectra are reported.  The $E_2^\mathrm{high}$ phonon mode was chosen for
detailed peak profile analyses.  The spectral background was removed
following Shirley,\cite{Shirley} and Breit-Wigner type peak fitting\cite{Lughi} 
was employed to determine peak position and the width.

Synchrotron x-ray scattering experiments were carried out at beam
line  11-ID-B of the Advanced Photon Source (Argonne National Laboratory)
using an x-ray energy of 90.8\,keV ($\lambda \approx 0.1365$\,\AA\/) at
room temperature.  Use of high energy radiation enables data acquisition 
at high $Q$ (=4$\pi$sin$\theta$/$\lambda$) wave vector, which in turn 
improves the reliability of the Fourier transformation for obtaining 
the PDF $G(r)$.  In the present work, scattering data
with $Q_\mathrm{max}$  of 28\,\AA\/$^{-1}$ were utilized for
extracting PDFs. Sample powders were  loaded in Kapton tubes and
data were measured in the transmission mode using an amorphous silicon
image plate system (General Electric Healthcare). For each sample, 33 images
were taken with an exposure time of 16\,s per image. The program 
\textsc{Fit2d}\cite{Hammers} was used to convert images into the 
corresponding one-dimensional x-ray diffraction (XRD) pattern. The
average crystal  structures of these samples have been determined by
Rietveld method and reported  previously.\cite{Kim} For obtaining the PDF
from XRD data, the  program \textsc{PDFgetX2} was used.\cite{Qiu2}
First the measured scattering  intensities were corrected for sample
container background, Compton scattering, and Laue diffuse scattering.
Next the coherently scattered intensities $I$($Q$) were normalized in
absolute electron units to give total scattering structure  functions
$S$($Q$). Finally the reduced structure functions $F(Q)=Q[S(Q)-1]$  were
Fourier-transformed to produce the atomic PDFs $G(r)$. The refinements 
of \ZMG\/ structures were performed against the above obtained
experimental PDFs using the software \textsc{PDFfit}.\cite{Proffen} 

\section{RESULTS}

\subsection{Sample characterization} 

\begin{figure} 
\smallskip \centering \epsfig{file=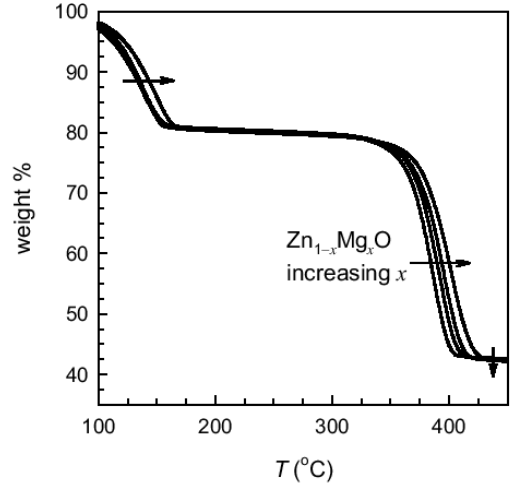, width=7cm}  
\caption{Thermogravimetry of \ZMG\/ ($x$ = 0, 0.05, 0.10, and 0.15) in air 
at a heating rate 5$^\circ$C\,min$^{-1}$.}
\label{fig:tga}
\end{figure}

The thermogravimetric profiles of
Zn$_{1-x}$Mg$_x$(C$_2$O$_4$)$\cdot$2H$_2$O  ($x$ = 0, 0.05, 0.10, and
0.15) are shown in Fig.\,\ref{fig:tga}. The stepwise  processes of
dehydration and oxalate-to-oxide conversion are clearly observed  at
around 130 and 390$^{\circ}$C, respectively,\cite{Dollimore} with the
weight  changes in good agreements with calculated estimates. With
increase of Mg content, the onset temperatures of both weight loss
steps shifted to higher temperature. This could be interpreted as indicative 
of the extra activation barrier for stabilizing Mg in the wurtzite lattice. 
For all cases, oxide formation was nearly complete at 500$^{\circ}$C
without any noticeable weight change at higher temperature.  It is
therefore assumed that isothermal heating at 550$^{\circ}$C for
24\,h  completely transforms Zn$_{1-x}$Mg$_x$(C$_2$O$_4$)$\cdot$2H$_2$O 
to the oxides \ZMG\/. 

\begin{figure}  
\smallskip \centering \epsfig{file=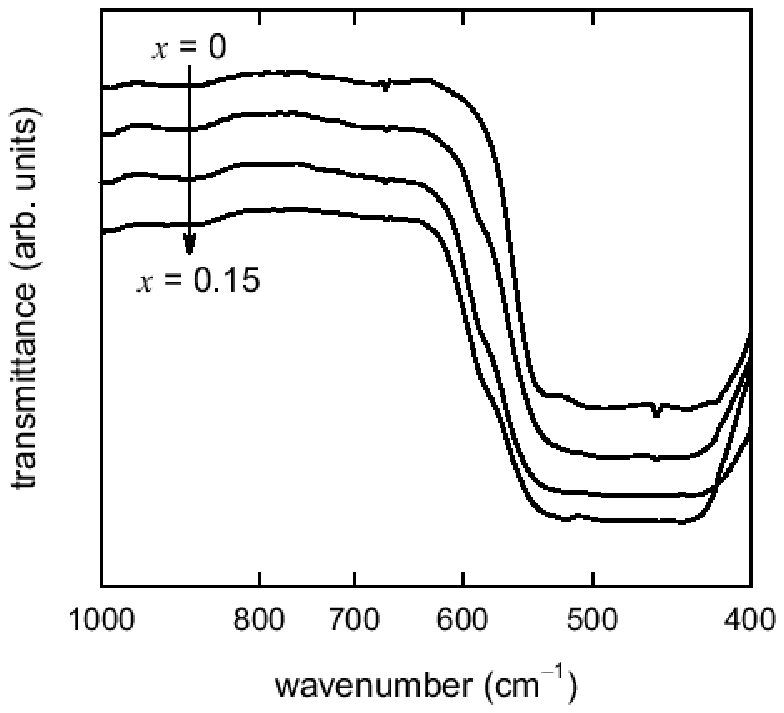, width=7cm} 
\caption{FT-IR spectra of \ZMG\/ ($x$ = 0, 0.05, 0.10, and 0.15).} 
\label{fig:ir}  
\end{figure}

In Fig.\,\ref{fig:ir}, the FT-IR spectra for \ZMG\/ are shown, which are
consistent with the registered reference data of wurtzite
ZnO.\cite{Smith}  A previous report suggested that organic precursors for
ZnO synthesis may leave  carbonate species strongly bound within the
lattice,\cite{Hlaing} but the samples  in this study did not exhibit any
spectral feature at $\approx$1300 and  $\approx$1500\,cm$^{-1}$
demonstrating the complete combustion of oxalate.\cite{Saussey}
The wurtzite lattice vibrations were observed as broad  IR bands at
400$-$600\,cm$^{-1}$. Upon Mg-substitution, these stretching modes
shifted to higher wavenumber as a result of the smaller reduced mass of 
Mg$-$O compared with that of Zn$-$O.

\begin{figure}  
\smallskip \centering \epsfig{file=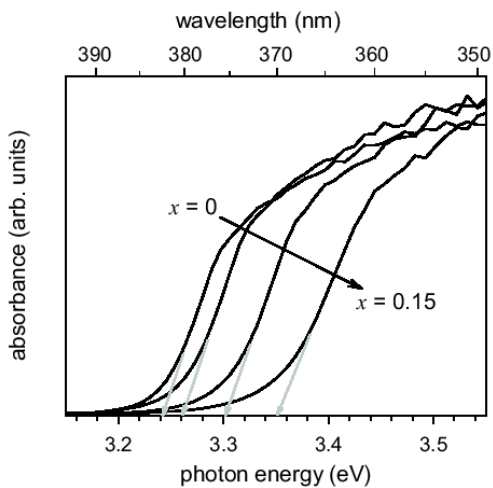, width=7cm} 
\caption{Diffuse-reflectance absorption spectra for \ZMG\/ ($x$ = 0,
0.05, 0.10, and  0.15). For each spectrum, the band gap energy is
indicated by the extrapolated arrow.} 
\label{fig:vis}
\end{figure} 

The optical band gaps ($E_\mathrm{g}$) of \ZMG\/ powders were determined from
the diffuse-reflectance absorption spectra as shown in
Fig.\,\ref{fig:vis}. The absorbance ($A$) of the samples were obtained
from the measured reflectance ($R$) according to the Kubelka-Munk
relation.\cite{Kubelka} 
\begin{equation} A = \frac{(1-R)^2}{2R}
\end{equation}  
In the present measurement setup, the sample analytes are
thick enough ($\approx$1.5\,mm) to disallow transmission, and the
Kubelka-Munk theory is appropriate.  As clearly seen in
Fig.\,\ref{fig:vis}, the absorption edges are shifted to the higher 
energy side with the increase of Mg content $x$. Also noticed are 
slight broadenings  of the edge slopes upon the Mg-substitution. The band
gap energy gradually increased from  3.24 ($x$ = 0) to 3.26 ($x$ = 0.05),
3.30 ($x$ = 0.10), and 3.35\,eV ($x$ = 0.15).  In previous studies on
\ZMG\/ thin films grown by pulsed laser deposition, Ohtomo $et\,al$. 
have observed a similar composition-dependence of the band gaps;
monotonic increases from  3.30 ($x$ = 0) to 3.63\,eV ($x$ =
0.14).\cite{Ohtomo} Their absolute band gap energies  are not exactly
reproduced in our polycrystalline samples, but this may be due  to
the differences in experimental details. The optical band
gaps of  solids are frequently determined to be different depending on sample
type, optical  characterization technique, and band gap determination
method.\cite{Srikant,Singh}  In the above study on \ZMG\/ films the band
gaps were evaluated with the assumption  $\alpha^2 \propto
(h\nu-E_\mathrm{g})$, in which the absorption coefficient $\alpha$  was
deduced from the transmittance measurements. 

\subsection{Raman spectroscopy}  

\begin{figure}  
\smallskip \centering \epsfig{file=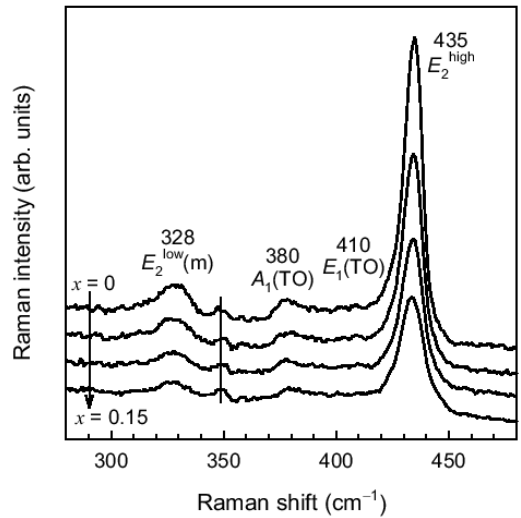, width=7cm} 
\caption{Raman spectra of the \ZMG\/
($x$ = 0, 0.05, 0.10, and 0.15). A plasma line  from the 488.08 nm
Ar$^{+}$ laser appears at $\approx$348\,cm$^{-1}$  (indicated by a
vertical line).} 
\label{fig:Raman1}
\end{figure} 

Wurtzite ZnO belongs to 
the space group $C_{6v}^{4}$ (Hermann-Mauguin symbol $P6_{3}mc$) and 
has six lattice phonon modes, $A_{1}+2B_{1}+E_{1}+2E_{2}$, of which the
$B_1$ branches  are Raman inactive.\cite{Calleja,Damen} The $A_1$ and 
$E_1$ modes are polarized along the $z$-direction and $xy$-plane,
respectively,  whereas the two $E_2$ modes ($E_2^\mathrm{low}$ and
$E_2^\mathrm{high}$) are non-polar.\cite{Calleja,Damen,Zhang} Each of
the polar modes is split to longitudinal (LO) and transverse optical 
(TO) components due to the macroscopic electric field associated with
the LO phonons.

\begin{figure}
\smallskip \centering \epsfig{file=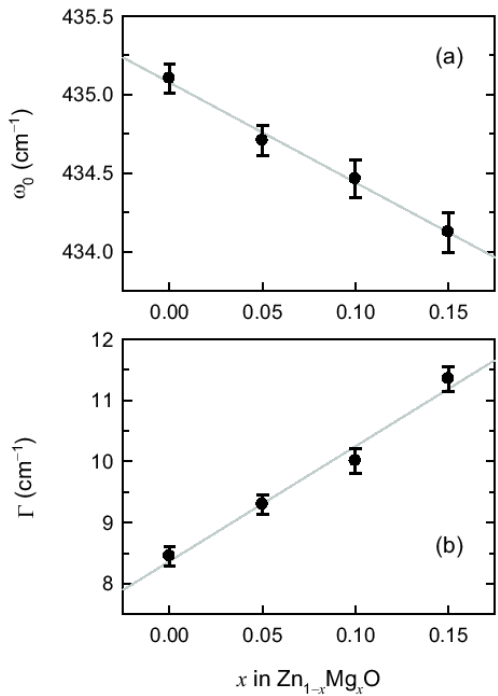, width=7cm} 
\caption{(a) phonon frequency $\omega_0$ and (b) width $\Gamma$ of 
Raman $E_2^\mathrm{high}$ mode as functions of Mg content in \ZMG\/. 
Gray lines are best linear fits.}
\label{fig:Raman2}
\end{figure}

Raman spectra of polycrystalline \ZMG\/ powders were examined over the
frequency  range 90$-$830\,cm$^{-1}$ and part of those spectra are
presented in  Fig.\,\ref{fig:Raman1}. For all cases of $x = 0-0.15$, the
measured Raman spectra agree well with the wurtzite ZnO vibration modes,
without any new bands arising from Mg-substitution. Raman features
characteristic of cubic MgO\cite{Ishikawa} were not observed.
In previous Raman studies on ZnO films or nanophases doped with N,
Al, Ga,  Sb, Fe, Mn, or Mg, additional bands have been observed and
attributed to the  induced lattice defect or the dopants' local
vibration.\cite{Bundes,Kaschner,Pan} A previous Raman study of 
polycrystalline \ZMG\/ has shown that the Mg-substitution flattens 
the overall Raman signal.\cite{Tomar} In Fig.\,\ref{fig:Raman1}, the 
Raman peaks from $A_1$(TO), $E_1$(TO), $E_2^\mathrm{low}$ (multiphonon), 
and $E_2^\mathrm{high}$ modes are found. The $A_1$ and $E_1$ modes 
reflect the strength of the polar lattice bonds, which are of interest 
in relation with the $c$-axial displacements of cations. However the 
peaks corresponding to those two modes  were not well resolved from 
the background and could not be used for a detailed study. Usually 
Raman signals from polar modes are weaker in intensity due to 
phonon-plasmon interactions.\cite{Bergman} The $E_2^\mathrm{high}$ 
phonon, which is the most prominent in the ZnO Raman spectra, was used 
for the peak profile  analysis. The Raman line shape was fitted using 
the Breit-Wigner function, 

\begin{equation}
I(\omega) \propto \frac{[1 + 2\beta(\omega-\omega_0)/\Gamma]^2}
                       {1 + [2(\omega-\omega_0)/\Gamma]^2}
\end{equation}

\noindent where $I(\omega)$ is the Raman intensity at a given 
frequency, $\omega_0$ is the Raman shift, $\Gamma$ is the broadening
expressed as full-width at half-maximum (FWHM), and $\beta$ is an 
asymmetry parameter ($\beta$ = 0 for the symmetrical peak).\cite{Lughi} 
For each Raman spectrum of \ZMG\/, the $E_2^\mathrm{high}$ phonon 
frequency $\omega_0$ was determined along with $\Gamma$, as plotted 
in Fig.\,\ref{fig:Raman2}. 

If we consider the mass change in the
Mg-substitution for Zn, the phonon frequency  is expected to increase for
the samples with higher $x$. But the observed trend is contrary to this:
The $E_2^\mathrm{high}$ phonon frequency decreases upon the 
Mg-substitution. To explain this, it should be recalled that the
$E_2^\mathrm{high}$  mode of ZnO corresponds mostly ($\approx$85\%) to
the vibration of oxygen  atoms,\cite{Serrano} and is insensitive to the
mass substitution on the cation site.  On the other hand, since the
$E_2^\mathrm{high}$ mode is associated with atomic motions  on the
$xy$-plane, its phonon energy depends on the in-plane lattice
dimensions. In the \ZMG\/ solid solution, the hexagonal parameter $a$
monotonically increases  with $x$, hence softening the
$E_2^\mathrm{high}$ phonon mode. The decrease of  $E_2^\mathrm{high}$
mode frequency has been similarly observed from ZnO thin films  under the
tensile strain along the $xy$-plane.\cite{Gruber} 

In addition to the frequency shift, it is also noted that the 
$E_2^\mathrm{high}$ peaks are markedly broadened upon the 
Mg-substitution. The Raman mode frequency and the  peak width can be 
influenced by the random substitution of Zn with Mg through the 
compositional disorder effect.\cite{Richter,Tiong,Samanta} As inferred from 
the evolution of average crystal structures in \ZMG, the Mg atoms do not
have an identical  bonding geometry to that of Zn. In other words the
Mg-substitution alters the  translational symmetry of the wurtzite
lattice to modify the phonon oscillation field.  In previous Raman
studies on the alloy systems, the spatial correlation has been regarded
an adequate mechanism responsible for the red-shift of the mode
frequency  as well as the peak broadening.\cite{Richter,Tiong,Samanta} 
The model is based on the wave  vector uncertainty, $\Delta q=2\pi/L$, 
where $q$ is the phonon wave vector range,  $L$ is the length scale of 
phonon confinement, and $q$ has the phonon dispersion  relation 
$\omega(q)$ with the Raman frequency. It predicts that, as the $L$ becomes 
smaller, $\Delta q$ becomes larger so that a wider range of frequencies 
is allowed  for Raman scattering. 

In \ZMG\/ compounds as well, the presence of the hetero component Mg seems to
perturb the  wurtzite phonon modes. As shown in Fig.\,\ref{fig:Raman2},
both the frequency shift  and the peak broadening become greater for the
samples with higher Mg-concentration.  However, the symmetry of
$E_2^\mathrm{high}$ peak shape did not show significant  dependences on
the Mg-concentration, which is somewhat unexpected. This is in contrast to 
the spatial correlation model and also with the experimental examples of
disordered semiconductor crystals, where the Raman line shapes become more 
asymmetric as the peak widths increase.\cite{Richter,Tiong} 

\subsection{Pair distribution functions} 

\begin{figure} \smallskip
\centering  \epsfig{file=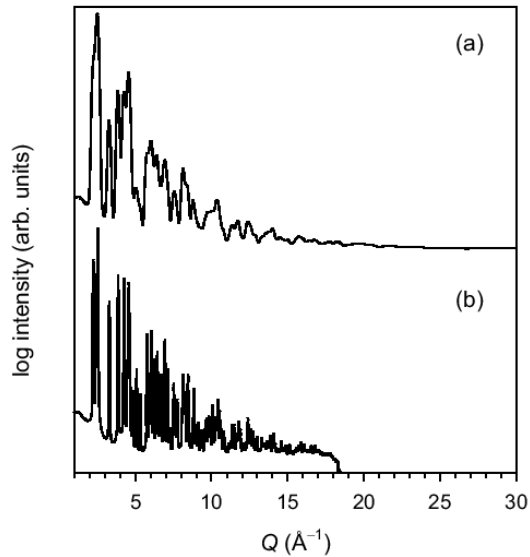, width=7cm}  \caption{XRD patterns of
Zn$_{0.95}$Mg$_{0.05}$O, used for (a) PDF analysis and  (b) Rietveld
refinement. Sample-to-detector distances were 150\,mm for (a) and
660\,mm  for (b).} \label{fig:xrd}
\end{figure}

Both Rietveld refinement and PDF analysis utilize one-dimensional 
diffraction patterns to extract crystal structure information. The former 
focuses on the long range average structure based on Bragg diffraction, 
while the latter focuses on the local structure information over short length  
scales from the diffuse scattering in addition to the Bragg 
diffraction.\cite{Bill,Qiu,Welber,Frey,Egami} 
Therefore the above two methods rely on distinct data qualities. 
The diffuse scattering becomes more prominent at high $Q$, for which 
the PDF data require very large Ewald spheres, but not necessarily very 
high resolution, and extremeley large signal to noise ratios. 
For Rietveld analyses, highly resolved Bragg peaks are preferred, at 
the expense of collecting data over smaller Ewald spheres. In order to 
meet both needs, the diffraction experiments were carried out at two 
different sample-to-detector distances, 150\,mm for the PDF data and 660\,mm 
for the Rietveld data. Fig.\,\ref{fig:xrd} compares the two XRD patterns of 
Zn$_{0.95}$Mg$_{0.05}$O, one used for Rietveld refinement and the other
for PDF study. The pattern for Rietveld refinement comprised sharp and 
well-resolved Bragg peaks  but the diffraction information rapidly
vanishes at $Q$ beyond 18\,\AA\/$^{-1}$. However, in the pattern for PDF
analysis the scattering intensities are observable at $Q$ as high as
30\,\AA\/$^{-1}$.  

\begin{figure}  \smallskip \centering \epsfig{file=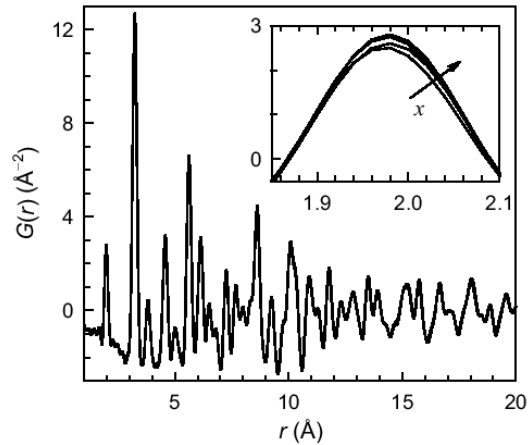, width=7cm} 
\caption{Experimental PDFs for \ZMG\/ phases with $x$ = 0, 0.05, 0.10,
and 0.15.  Inset shows the zoomed-in view of the peak corresponding to
the first coordination (Zn,Mg)$-$O shell.} \label{fig:pdfwide} 
\end{figure} 

The PDF data $G$($r$) for the four samples of \ZMG\/ ($x$ =
0, 0.05, 0.10, and 0.15) were obtained following a standard data 
processing sequence.\cite{Qiu2,Chupas}  In Fig.\,\ref{fig:pdfwide}, the
PDFs from different \ZMG\/ compositions are  superimposed, and it is observed 
that the data are hardly distinguished. Since the  average crystal
structures (lattice symmetry, lattice parameters, atomic positions, 
$etc$.) of \ZMG\/ phases are similar, the composition-dependent changes
in PDFs are not immediately noticeable. However a close view at the PDF
peak at $r\approx$ 2\,\AA\/ exposes a subtle evolution in the coordination
shell of the nearest cation$-$anion  bond pairs. As the Mg content
increases, these PDF peaks are slightly  shifted to longer $r$, with
gradual increases of peak heights. This finding  is consistent with the
previous Rietveld refinement result which indicated that the (Zn,Mg)$-$O
bond distances become more regular upon Mg-substitution.\cite{Kim}
The first shell interatomic distances are (3$\times$)1.9731\,\AA\/ and 
1.9941\,\AA\/ in ZnO, and (3$\times$)1.9756\,\AA\/ and 1.9873\,\AA\/  in
Zn$_{0.85}$Mg$_{0.15}$O. Although small in magnitudes, the above changes 
correspond to $\approx$45\% reduction in the standard deviation of the 
four bond distances, which result in the discernible peak sharpening.
The shifts of  peak positions can be ascribed to the lengthening of the
three shorter bonds rather  than a change in the longer apical bond.  

The real space PDF data are therefore useful for directly depicting the
bonding  geometry in a specific $r$-range. For more detailed and
quantitative information,  the PDF data were analyzed by full profile
refinements using several different models of simple cell as well as 
supercell. The simple cell models were based on the 
hexagonal wurtzite structure (space group $P6_{3}mc$, $Z$ = 2, 
$a\approx$ 3.25\,\AA\/, $c\approx$ 5.20\,\AA\/) with O at 
($\frac{1}{3}$\,$\frac{2}{3}$\,0) and Zn/Mg at
($\frac{1}{3}$\,$\frac{2}{3}$\,$u$). In the analyses using the simple
cell, the fractional occupancies of Zn and Mg  were set according to the
composition of each phase \ZMG\/ ($x$ = 0, 0.05, 0.10, and 0.15).
Supercell model structures were constructed by expanding the wurtzite 
simple cell to a $3\times 3\times 1$ size ($Z$ = 18, $a\approx$
9.75\,\AA\/,  $c\approx$ 5.20\,\AA\/). By partially replacing the Zn
atoms with Mg, the parent Zn$_{18}$O$_{18}$ supercell can be modified
to the compositions  Zn$_{17}$MgO$_{18}$ (Zn$_{0.944}$Mg$_{0.056}$O),
Zn$_{16}$Mg$_2$O$_{18}$  (Zn$_{0.889}$Mg$_{0.111}$O), and
Zn$_{15}$Mg$_3$O$_{18}$ (Zn$_{0.833}$Mg$_{0.167}$O)  to model the $x$ =
0.05, 0.10, and 0.15 phases, respectively. The Mg atoms in 
Zn$_{16}$Mg$_2$O$_{18}$ and Zn$_{15}$Mg$_3$O$_{18}$ models were separated
to be  as distant from one another as allowed by the cell. PDF
refinements\cite{Proffen,Egami} of $G(r)$ were carried out using the
variables of scale factor, dynamic correlation factor, resolution
factor, lattice constants, independent $u$ of Zn and Mg, and isotropic 
temperature factors of Zn, Mg, and O.  

\begin{table}
\caption{Summary of PDF refinements for \ZMG\/ ($x$ = 0, 0.05, 0.10, 
and 0.15) using simple cell models, over the $r$-range of 1.5$-$10\,\AA.} 
\begin{ruledtabular} 
\begin{tabular}{ccccc} 
$x$        & 0          & 0.05        & 0.10       & 0.15       \\
\hline
$R_\mathrm{w}$ (\%) & 13.4  & 17.0    & 17.2       & 17.8       \\
$a$ (\AA)  & 3.2500(2)  & 3.25028(5)  & 3.2504(1)	& 3.2506(1)  \\
$c$ (\AA)  & 5.2101(4)  & 5.2109(1)   & 5.2115(3)  & 5.2119(3)  \\
$u$(Mg)    & -          & 0.380(4)    & 0.379(4)   & 0.380(3)   \\
$u$(Zn)    & 0.3819(4)  & 0.3817(1)   & 0.3815(3)  & 0.3813(3)  \\
\end{tabular}
\end{ruledtabular} 
\label{tab:pdffit} 
\end{table}

\begin{figure}  \smallskip \centering \epsfig{file=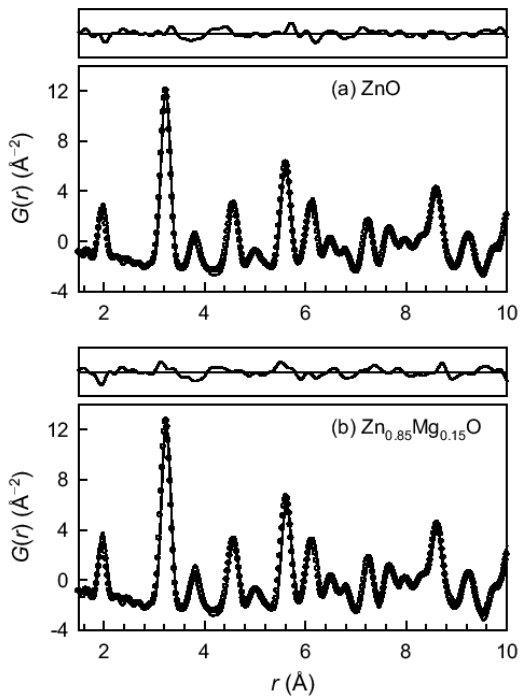, width=7cm} 
\caption{PDF refinement profiles for (a) ZnO and (b)
Zn$_{0.85}$Mg$_{0.15}$O,  using four-atom simple cell models.
Experimental (open circles) and calculated  (solid lines) data are
plotted in the bottom panels, along with the difference  patterns
($G_\mathrm{expt}-G_\mathrm{calc}$) in the top panels.}
\label{fig:pdffit}  \end{figure}

\begin{figure}  \smallskip \centering \epsfig{file=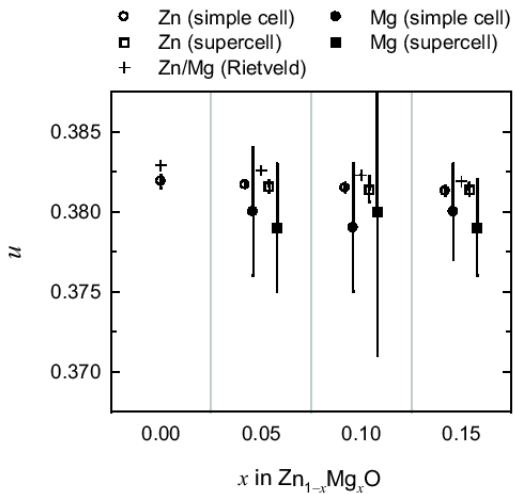, width=7cm} 
\caption{Position parameters for Mg and Zn in \ZMG\/ phases, as analyzed
by PDF refinements (1.5$-$10\,\AA\/) of simple cell and $3\times 3\times 1$ 
supercell model structures. Rietveld refinement results\cite{Kim} 
are also compared. For the PDF results, error bars indicate one 
estimated-standard-deviation (esd) while for the Reietveld results, 
the symbol height corresponds to one esd.}
\label{fig:pdfu}  \end{figure}

First the PDF refinements were performed for \ZMG\/ ($x$ = 0, 0.05, 0.10,
and 0.15)  using the four-atom simple cell models, in which the cation
compositions were  accounted for with fixed occupancy parameters.
Table\,\ref{tab:pdffit} summarizes the results of the 1.5$-$10\,\AA\/
range fitting, and Fig.\,\ref{fig:pdffit} shows the refinement profiles
for ZnO and Zn$_{0.85}$Mg$_{0.15}$O. Refinements over wider $r$-ranges
tended to yield higher $R_\mathrm{w}$ factors but the fit results were
similar to those given in Table\,\ref{tab:pdffit}. It can be mentioned
that the obtained $R_\mathrm{w}$'s near 20\% are common for PDF 
refinements, even for well-crystallized materials. The $R_\mathrm{w}$'s
from  the PDF and Rietveld refinements cannot be directly compared, but
they serve the same purpose of finding out the best structure solution
from a number of competing models with similar numbers of refinable
parameters.\cite{Egami,Petkov} The inherently higher
$R_\mathrm{w}$ in the PDF analysis stems from the greater sensitivity 
to the local atomic ordering, imperfect data correction,
and systematic errors.\cite{Egami}  The $R_\mathrm{w}$'s for the \ZMG\/
($x > 0$) samples are significantly higher than that for ZnO ($x = 0$).
It implies the presence of irregular Mg/Zn distribution in the alloy
phases, which may not be well portrayed by simple structure models. 

For all the four compositions, the lattice constants from the PDF agree 
well with the Rietveld refinement results, although the 
precision is slightly lower in PDF. Moreover the Zn atomic 
position and the Zn$-$O bond distances in ZnO are well reproduced in the
PDF analysis, when compared with Rietveld refinement results. For
the \ZMG\/ phases, the $z$-coordinates of Zn and Mg were independently
refined to investigate the distinct local geometries of the two cation
types. Interestingly the refined $u$(Mg) is smaller than $u$(Zn) for all 
the three solid solutions with $x$ = 0.05, 0.10, and 0.15 
(Fig.\,\ref{fig:pdfu}). This trend is consistently repeated in the supercell 
model analyses detailed below. The $u$(Mg) parameters have rather large 
estimated deviations together with weak variations depending on $x$, but
the observation $u$(Mg) $<$ $u$(Zn) was quite robust regardless of the
refinement $r$-range or the choice of constraints.  

\begin{figure*} 
\smallskip \centering \epsfig{file=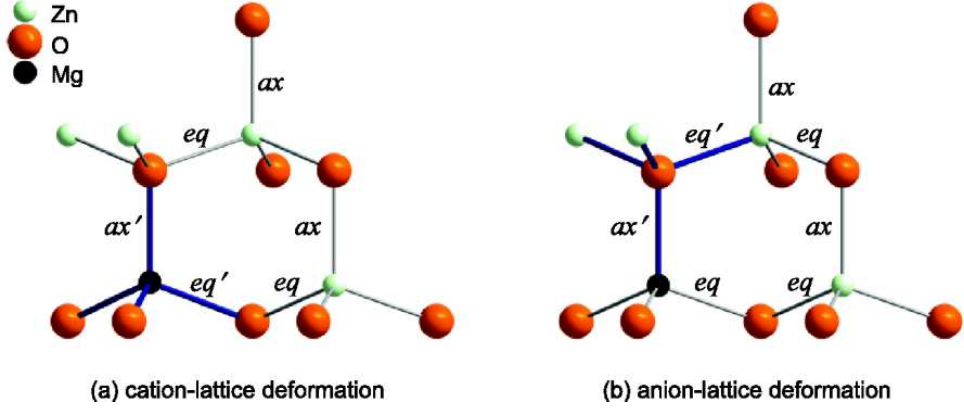, width=13cm} 
\caption{(Color online) Local structural distortions in \ZMG\/ supercells 
of (a) cation-lattice  deformation, $u$(Mg) $\neq$ $u$(Zn), $u$(O$_\mathrm{Mg}$)
= $u$(O$_\mathrm{Zn}$)  and (b) anion-lattice deformation, $u$(Mg) =
$u$(Zn), $u$(O$_\mathrm{Mg}$) $\neq$  $u$(O$_\mathrm{Zn}$). $ax$ and $eq$
denote the cation$-$anion distances with  different bond orientations,
$\parallel c$ and $\perp c$, respectively.  As resulted from local
distortions, $ax^\prime$ $\neq$ $ax$ and $eq^\prime$ $\neq$ $eq$.}
\label{fig:tdstr} 
\end{figure*} 

We recognize that in the above simple cell model for \ZMG, only the cation 
sub-lattices can be distorted while the anion framework remains undistorted. 
The wurtzite structure is its own antitype where the cation and anion 
sublattices are interchangeable. Depending on which of the two is distorted the
resulting PDFs will be slightly different from each other, although the
corresponding average crystal structures are practically
indistinguishable.  In case the oxygen-lattice is distorted, the partial
PDF from O$-$O pairs is  modified from that of the average crystal
structure, and likewise, distortion  of the cation-lattice affects the
PDF for (Zn,Mg)$-$(Zn,Mg) pairs.  The differentiations of cation$-$anion
bond lengths in \ZMG, resulting from  those two deformation cases, are
illustrated in Fig.\,\ref{fig:tdstr}. In order to  verify the lattice
deformation type, we have examined the PDF refinements of 
Zn$_{0.85}$Mg$_{0.15}$O using two different supercell models of 
Zn$_{15}$Mg$_3$O$_{18}$. In the anion-lattice deformation model, the
$z$-coordinates  of Zn and Mg atoms were fixed to 0 or $\frac{1}{2}$, but
the $z$-coordinates of  O atoms were refined with constraints; the three
O atoms (O$_\mathrm{Mg}$) that are  apical to Mg were bound to have
symmetry-related $z$-coordinates  ($z_j$ = $z_i$ or $z_{i}+\frac{1}{2}$),
and so were the other fifteen O atoms  (O$_\mathrm{Zn}$). In the
cation-lattice deformation model, the $z$-coordinates of  O atoms were
fixed, while the Zn and Mg positions were allowed to vary in independent 
groups. For each model, PDF refinements were carried out for the
$r$-ranges from  1.5\,\AA\/ to various $r_\mathrm{max}$ of
6.4$-$­21.4\,\AA. Attempts using the smaller  $r_\mathrm{max}$ ($<$
6\,\AA) suffered instability of the lattice constants, and  were
discarded. Since the resolution factor was very sensitive to the
$r_\mathrm{max}$,  it was fixed to 0.1 in all the refinements.

\begin{figure}  \smallskip \centering \epsfig{file=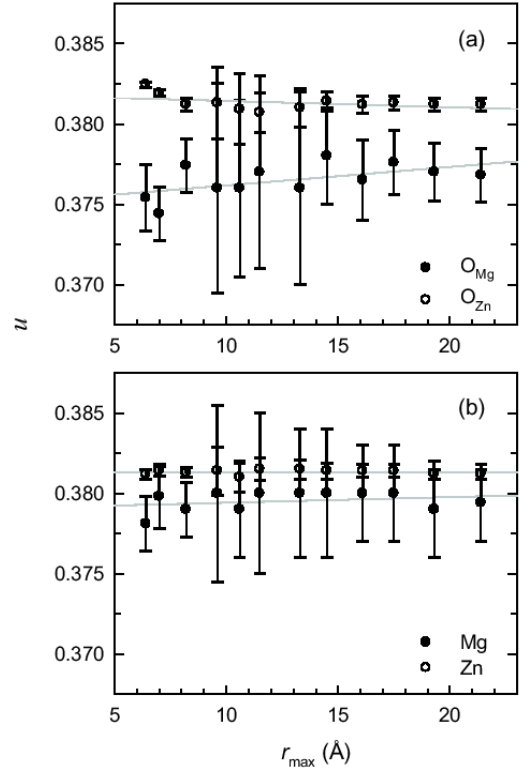, width=7cm} 
\caption{Atomic positions in Zn$_{0.85}$Mg$_{0.15}$O as functions of the
length of  the vector $r_\mathrm{max}$. PDF refinements were carried out
using the supercell model  structures based on (a) anion-lattice
deformation and (b) cation-lattice deformation. Gray lines are best linear 
fits.} \label{fig:pdfsuperu} 
\end{figure}

The refinement results using the two supercell models are presented 
in  Fig.\,\ref{fig:pdfsuperu}. For both models, the $R_\mathrm{w}$  factors 
were obtained in the range of 17$\sim$19.5\% with no apparent 
$r_\mathrm{max}$-dependences. The two models are based on the deformation
of different sublattices, and accordingly the $z$-coordinates of either
cation or anion are presented for each. However, it can be well assumed
that $u$(Mg) and $u$(Zn)  from the cation-lattice deformation model are
comparable respectively to  $u$(O$_\mathrm{Mg}$) and $u$(O$_\mathrm{Zn}$)
from the anion-lattice deformation.  As shown in
Fig.\,\ref{fig:pdfsuperu}, the two supercell models similarly indicate 
that $u$(Mg) $<$ $u$(Zn) and $u$(O$_\mathrm{Mg}$) $<$
$u$(O$_\mathrm{Zn}$), namely  that the axial Mg$-$O bonds are shorter
than the axial Zn$-$O bonds. A quick  examination of
Fig.\,\ref{fig:pdfsuperu} reveals that the cation-lattice  deformation
model provides more reproducible refinements of atomic positions.  The
averages over the $r_\mathrm{max}$-dependent refinement results are
$u$(Mg)  = 0.3795(6) vs. $u$(O$_\mathrm{Mg}$) = 0.3765(10) and $u$(Zn) =
0.3813(1)  vs. $u$(O$_\mathrm{Zn}$) = 0.3813(5). The comparison of
lattice constants (not shown) further showed that the anion-lattice 
deformation model yields more scattered results. It is therefore judged 
that the cation-lattice  deformation is appropriate for
analyzing the \ZMG\/ structure. 

\begin{figure}  \smallskip \centering \epsfig{file=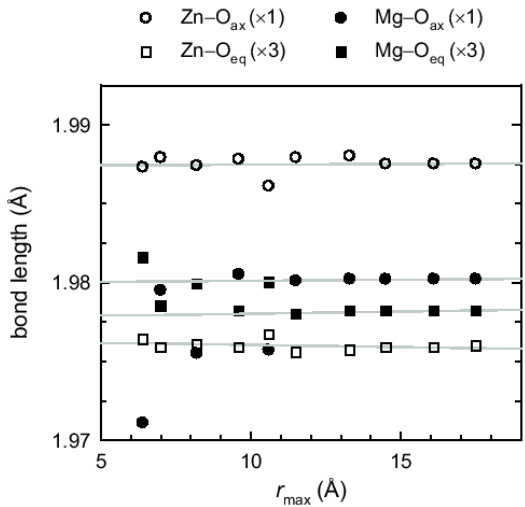, width=7cm} 
\caption{Zn$-$O and Mg$-$O bond distances in Zn$_{0.85}$Mg$_{0.15}$O,
obtained from the PDF refinements using the cation-lattice deformation
supercell model. Gray lines are guides to the eye.} 
\label{fig:pdfbd} 
\end{figure}

Here it can be instructive to compare the local distortions in both
deformation  models in detail. In the cation-lattice deformation model
(Fig.\,\ref{fig:tdstr}a),  the bonding geometries of cations are
differentiated simply by the atom type,  Zn or Mg. On the other hand the
anion-lattice deformation (Fig.\,\ref{fig:tdstr}b)  creates three cation
groups having different bonding geometries; (i) three Mg atoms,  (ii)
nine Zn atoms that are bonded to O$_\mathrm{Mg}$, and (iii) six Zn atoms
that are  not bonded to O$_\mathrm{Mg}$. For each cation group, the four
nearest anion distances  are slightly different, (i) $ax'+3eq$, (ii)
$ax+2eq+eq'$, and (iii) $ax+3eq$, using the notations in Fig.\,\ref{fig:tdstr}. 
However, the groups (i) and (iii)  have same bond angles about the cation. 
It might be possible that the Zn atoms  (ii and iii) have non-uniform 
coordination geometries depending on the proximity to the Mg, but it is 
hardly expected that Zn (iii) and Mg (i) have the same bond angles. 
In this regard, the coordination geometries of Zn and Mg are not sensibly 
distinguished in the anion-lattice deformation model. 
In Fig.\,\ref{fig:pdfbd}, the Zn$-$O and Mg$-$O bond distances
in Zn$_{0.85}$Mg$_{0.15}$O are plotted as obtained from the
cation-lattice deformation model. The refined structural parameters ($a,
c, u$) were used to calculate the bond distances. For both Zn and Mg,
the first coordination shells consist of one longer ($c$-axial) bond and 
three shorter (equatorial) bonds. As reflected by the difference between 
the longer and shorter bonds, Mg atoms are found to sit in more regular
tetrahedra of oxygen than are Zn atoms. 

\section{DISCUSSION}

For various $AX$-type compounds including ZnO and MgO, the relative
stabilities  of wurtzite, zinc blende, and rock salt structures have been
a long-standing and  intriguing subject.\cite{Phil,Mooser,OKeeffe} The
correlations between the binary  composition and the favored structure
type have been proposed in several different  ways, but commonly using
the bond ionicity and the ion sizes of $A$ and $X$ as  the principal
parameters.\cite{Phil,Mooser,OKeeffe} If we limit our attention to  ZnO
and MgO, the observed crystal structures can be explained simply by the 
ionicity consideration, in other words, electronegativities of Zn and
Mg.  With intermediate electronegativity, Zn can achieve the
tetrahedrally directed  valence orbital \textit{via} facile 
$sp^3$-hybridization. On the other hand,  Mg is  more electropositive
and its valence orbital is dominated equally by the  3$p_x$, 3$p_y$, and
3$p_z$ components, by which the nearest oxide ions are oriented  in the
octahedral geometry. Hence Zn and Mg adopt the four- and six-fold 
coordinations, respectively, in their binary oxides. Given those
coordination preferences, the alloying with MgO should destabilize the
host ZnO lattice. Indeed the thermodynamic solubility limit in \ZMG\/ is
reached at a low Mg-concentration, $\approx$15\%.\cite{Kim,Sapozh} 

The ZnO has wurtzite parameters of $c/a$ = 1.6021 and $u$ = 0.3829,
which are ideally 1.633 and 0.375, respectively.\cite{Kim} Considering
the range of $c/a$ (1.600$\sim$1.645) observed from the binary
wurtzites,\cite{Lawaetz} ZnO is  near the borderline of wurtzite
stability. On the other hand, the wurtzite MgO structure is not
available experimentally. Instead, the first-principle methods based on
density functional theory (DFT) have predicted that the wurtzite MgO
will have much smaller $c/a$ than those of existing wurtzite phases. 
An earlier DFT study has reported an extremely distorted hexagonal
structure ($c/a$ = 1.20, $u$ = 0.5) for the wurtzite MgO, in which Mg
sits on a mirror plane and has a bi-pyramidal coordination.\cite{Limp}
However recent studies, independently conducted by Janotti
$et\,al$.\cite{Janotti} and Gopal and  Spaldin,\cite{Gopal} have reported
$c/a$ = 1.51 and $u$ = 0.398. More relevant to our experimental work,
Malashevich and Vanderbilt performed the geometry optimization of \ZMG\/
supercells ($x$ = 0, $\frac{1}{6}$, $\frac{1}{4}$, $\frac{1}{3}$, and
$\frac{1}{2}$).\cite{Malash} They showed that the $c/a$  decreases with
increase of Mg content, and $u$(Mg) is larger than  $u$(Zn) in
all the supercell compositions. Therefore those DFT studies on MgO and 
\ZMG\/ consistently surmise that the Mg-substitution in ZnO would result
in the  decrease of $c/a$, in agreement with the previously established
idea that the  $c/a$ deviates farther from the ideal, when the bonding
character becomes more ionic.\cite{Schulz} Our Rietveld study on \ZMG\/
has also shown that $c/a$ is  gradually decreased with the
Mg-substitution.\cite{Kim} The Mg-substitution in ZnO is not disfavored
from packing considerations, but may reduce the electrostatic  stabilizations
in the long range through the change in $c/a$ ratio. At the  solubility
limit Zn$_{0.85}$Mg$_{0.15}$O, the $c/a$ is as low as 1.600, and  this
value seems to represent the lowest $c/a$ that can be sustained by the 
wurtzite lattice. 

The energy-minimized \ZMG\/ supercells of Malashevich and
Vanderbilt\cite{Malash2} have $u$(Mg) in the range 0.387$\sim$0.390,
and the $u$(Zn), 0.378$\sim$0.382, which is not in agreement with the 
PDF results shown in Figs.\,\ref{fig:pdfu} and \ref{fig:pdfsuperu}. As a
consequence of $u$(Mg) $>$ $u$(Zn), the spontaneous  polarization in the
computed \ZMG\/ supercells become larger for the compositions  with
higher Mg content,\cite{Malash} while the Rietveld analysis indicated
that the $u$ parameter of \ZMG\/ average structures and therefore the
spontaneous  polarization in the Clausius-Mosotti limit decrease upon 
Mg-substitution.\cite{Kim} The above
discrepancies are in  fact connected with a more fundamental question of
how the tetrahedral geometries of ZnO$_4$ and MgO$_4$ would differ in
the extended solids. Understanding of cation local geometries in
\ZMG\/ is important also for the application of ZnO/\ZMG\/
heterojunction devices. While tetrahedral MgO$_4$ is found in a
few minerals such as MgAl$_2$O$_4$, Ca$_2$MgSi$_2$O$_7$,\cite{Kimata}
and K$_2$MgSi$_5$O$_{12}$,\cite{Bell} \ZMG\/ provides an opportunity
to explore its equilibrated tetrahedral geometry. However
the computational and experimental results on the internal
tetrahedral geometry of \ZMG\/ do not agree well. In the supercell
structures of Malashevich and  Vanderbilt,\cite{Malash2} the ZnO$_4$
tetrahedra are more regular than the MgO$_4$ tetrahedra (as judged 
from the differences of the cation$-$anion bond distances), in contrast 
to the PDF findings presented in Fig.\,\ref{fig:pdfbd}. These 
points may be further clarified by using alternative techniques such as
neutron PDF or extended x-ray absorption fine structure. It is equally 
possible that the DFT methods must be reexamined, with inclusion of 
perhaps more accurate exchange functionals.

\section{CONCLUSION}

Raman and synchrotron x-ray PDF have been employed to probe the local
structures of  Zn and Mg atoms that occupy the common crystallographic
site in the wurtzite  alloys \ZMG\/ ($x$ = 0, 0.05, 0.10, and 0.15).
Regardless of the Mg concentration, structure model, and $r$-range, the
PDF refinements consistently show that the  Mg atoms have smaller
out-of-center tetrahedral displacements than the Zn atoms.  Even for the
Zn$_{0.95}$Mg$_{0.05}$O that has only 5\,mol\% MgO, the atomic 
coordinates of Zn and Mg were similarly obtained as those for 
Zn$_{0.90}$Mg$_{0.10}$O and Zn$_{0.85}$Mg$_{0.15}$O, thereby
demonstrating  the fidelity of PDF technique. It is understood that the
lessened tetrahedral  distortion of MgO$_4$, compared with ZnO$_4$, leads
to the decrease of wurtzite  $u$ parameters in the \ZMG\/ average
structures. The hexagonal $c/a$  ratio of \ZMG\/ decreases with the
increase of Mg concentration, which is ascribed  to the ionic character
contributed from MgO. Therefore, from both ionicity and  dipole strength
viewpoints, we expect that the \ZMG\/ alloys will have smaller 
spontaneous polarizations than ZnO. It is noted that in terms of 
ZnO$_4$ and MgO$_4$ local geometries, our experimental
results are somewhat contrary to the descriptions obtained from DFT-based
studies. 

\acknowledgments The authors gratefully acknowledge discussions with
Andrei Malashevich and David  Vanderbilt and support from the National
Science Foundation through the MRSEC  program (DMR05-20415). K.P. has
been supported by an NSF Graduate Student Fellowship. Work at Argonne
National Laboratory and the Advanced Photon Source was supported by
Department of Energy, Office of Basic Energy Sciences under Contract No.
W-31-109-Eng.-38. The authors thank Peter Chupas and Karena Chapman for
the help with synchrotron data collection.

\end{document}